**Interview with Adrian Raftery**

Leontine Alkema, Thomas Brendan Murphy and Adrian E. Raftery

**Abstract:** Professor Adrian E. Raftery is the Boeing International Professor of Statistics and Sociology, and an adjunct professor of Atmospheric Sciences, at the University of Washington in Seattle. He was born in Dublin, Ireland, and obtained a B.A. in Mathematics (1976) and an M.Sc. in Statistics and Operations Research (1977) at Trinity College Dublin. He obtained a doctorate in mathematical statistics in 1980 from the Université Pierre et Marie Curie in Paris, France under the supervision of Paul Deheuvels. He was a lecturer in statistics at Trinity College Dublin from 1980 to 1986, and then an associate (1986-1990) and full (1990-present) professor of statistics and sociology at the University of Washington. He was the founding Director of the Center for Statistics and Social Sciences (1999-2009).

Professor Raftery has published over 200 articles in peer-reviewed statistical, sociological and other journals. His research focuses on Bayesian model selection and Bayesian model averaging, model-based clustering, inference for deterministic simulation models, and the development of new statistical methods for demography, sociology, and the environmental and health sciences.

He is a member of the United States National Academy of Sciences, a Fellow of the American Academy of Arts and Sciences, an Honorary Member of the Royal Irish Academy, a member of the Washington State Academy of Sciences, a Fellow of the American Statistical Association, a Fellow of the Institute of Mathematical Statistics, and an elected Member of the Sociological Research Association. He has won the Population Association of America's Clifford C. Clogg Award, the American Sociological Association's Paul F. Lazarsfeld Award for Distinguished Contribution to Knowledge, the Jerome Sacks Award for Outstanding Cross-Disciplinary Research from the National Institute of Statistical Sciences, the Parzen Prize for Statistical Innovation, and the Science Foundation Ireland St. Patrick's Day Medal. He is also a former Coordinating and Applications Editor of the Journal of the American Statistical Association and a former Editor of Sociological Methodology. He was identified as the world's most cited researcher in mathematics for the decade 1995-2005 by Thomson-ISI.

Thirty-three students have obtained Ph.D.'s working under Raftery's supervision, of whom 21 hold or have held tenure-track university faculty positions. He has over 150 academic descendants.

This interview took place over two sessions in March 2023.

## Early years

### Do you want to tell us about your family background?

I was born in Dublin and one of four children. My oldest sister, Mary, died in 2012. She was very well known in Ireland, a documentary filmmaker who broke open the pedophilia scandals in the Catholic Church in Ireland in the late 1990s. I also have a younger twin brother and sister.

### When did your interest in statistics and other disciplines begin?

Well, statistics came first. In my family, we had more literary interests than quantitative ones, which is usually the case in Ireland. Quantitative subjects haven't been as prominent historically. But when I was 11, my father gave me two books for my birthday. One was M. J. Moroney's "Facts from Figures," an introduction to statistics, and the other was Darrell Huff's "How to Take a Chance," an introduction to probability (Figure 1). Even though I didn't fully understand them, I found them fascinating, and they hooked me on the idea of statistics. I even designed some card games based on roulette using the concepts from the books and made some money from them.

In terms of mathematics, I attended a small private school called St. Conleth's College in Dublin (Figure 1), where I had an inspirational mathematics teacher named Michael Manning. Several people from that school ended up pursuing mathematics. One notable person is Duncan Temple Lang, who is a faculty member at UC Davis in California. Another close friend, Maurice O'Reilly, became the president of the Irish Mathematical Society. So, mathematics was important despite it being a small school.

After graduating from the school at a young age (15), I couldn't go to college yet because I was too young. I spent a year working in an actuarial office, evaluating pension funds and providing advice on personal injury cases. During that time, I discovered discrepancies in the mortality rate schedule used by a shipping company based in Liverpool. I performed a chi-squared test that showed that the schedule didn't fit the data, but my senior colleagues just found this amusing and continued using the standard schedule. This experience sparked my interest in mortality and demographic studies. Throughout my college years and PhD, I also pursued actuarial exams. By the time I completed my PhD, I had an actuarial qualification. However, I chose the academic path over a higher-paying actuarial position, which turned out to be a fortunate decision.

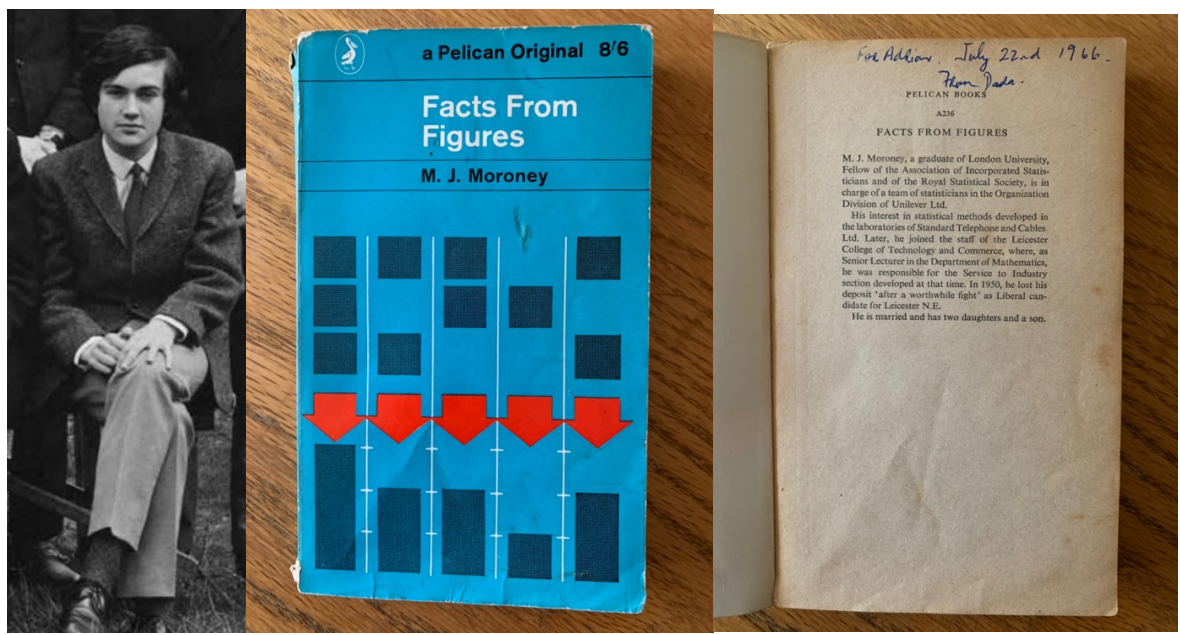

Figure 1. Left: Adrian at St. Conleths College in 1971. Center and right: The book that Adrian's father gave him for his birthday in 1966 that sparked his interest in statistics.

### Before we get to your academic path after your PhD, can you tell us about your undergraduate and graduate studies?

I went to Trinity College Dublin, where I did a mathematics degree. The program focused solely on mathematics courses for four years, unlike the broader American system. I then completed a master's degree in statistics and operations research, working on a thesis on solar energy under the supervision of John Haslett. My thesis, which was a group project with Philip Shier and Titi Obilade, involved analyzing a deterministic simulation model for solar energy to estimate the potential energy generation in a typical Irish house. Ireland is known for its rain and clouds, so at the time it was assumed that solar energy wouldn't be viable. However, we found out that over half of radiation penetrates through clouds, making solar energy a feasible option. This project combined deterministic modeling with statistical analysis, which became a significant theme in my research. After completing my masters degree, I pursued a PhD in Paris.

### Why did you choose to study in Paris?

My father was an Irish diplomat; and when I was 11, he was assigned to the Irish Embassy in Paris. So, our family went to live there, and I was thrown into the French school and learned French, and I liked that. I wanted to go back to Paris. Then the French government, when it came time, gave me a fellowship to do this.

Initially, I joined the probability department at the Université Paris 6 (now Sorbonne-Université), hoping to work on stochastic modeling of social processes inspired by the work of David Bartholomew in London. However, the focus in Paris was primarily on general theory and mathematical aspects of stochastic processes such as Brownian motion and Gaussian processes, so I learned a lot about that from Jacques Neveu, Klaus Krickeberg and Marc Yor. After a year, I switched to statistics, which was also mathematically oriented. Although I learned a great deal, it wasn't exactly what I had in mind. My PhD advisor, Paul Deheuvels, was very supportive, but he and I had different interests, and while we maintained good contacts later, we didn't collaborate. My PhD thesis was on non-Gaussian time series and asymptotic properties of estimators.

## Early Career

After your PhD, you started at Trinity College Dublin in the statistics department as lecturer or Assistant professor. How were those years, what were some highlights?

My time on the faculty at Trinity College Dublin was actually a very important stage of my career (Figure 2). I was there for five years, 1980-1985. It was a great environment for interaction. The department had about 10 or 11 people, and we would all have coffee together in the morning at eleven, then we'd have lunch together, and then we'd have tea together in the afternoon at four. It was a terrific atmosphere for interaction, and I learned a great deal from the discussions about research and statistics. Looking back, I realize that many research strands started in Trinity and my interdisciplinary work started in Trinity as well. It was a very fruitful period.

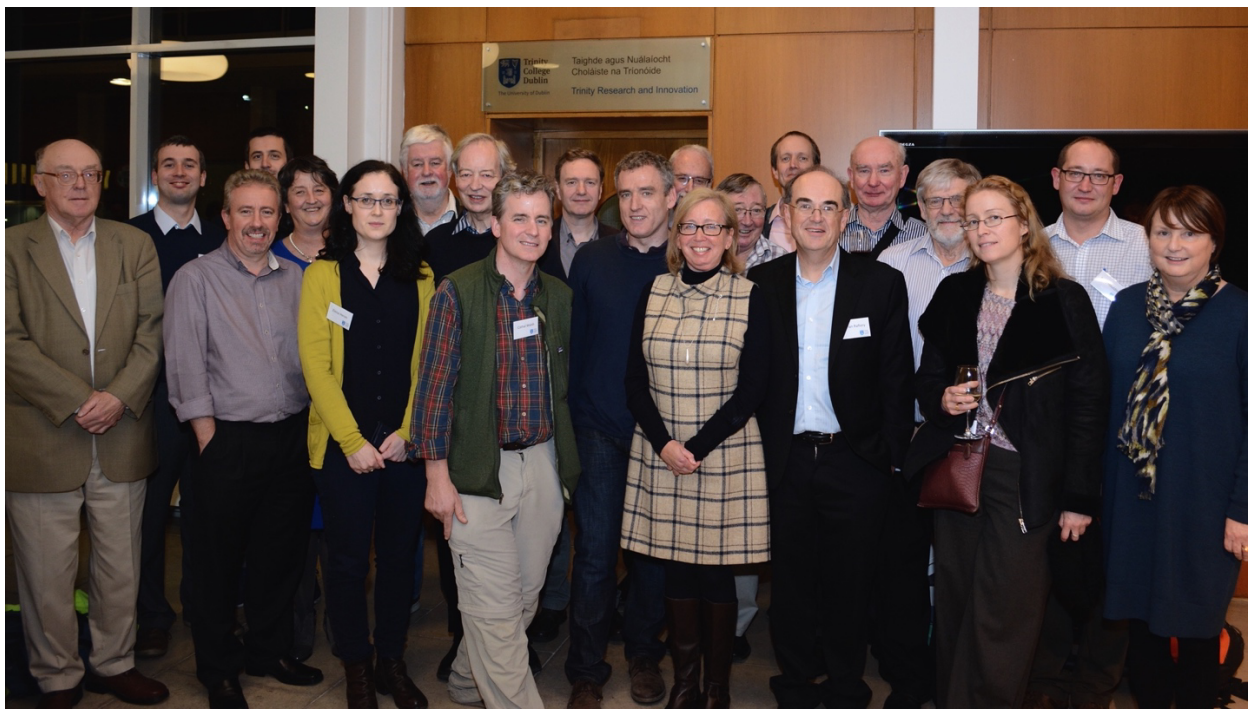

Figure 2: Past and present members of the Trinity College Dublin Statistics Department at the 50th anniversary in 2017. Left to right: Tony Kinsella, Arthur White, John McGill, Bernardo Nipoti, Myra O'Regan (colleague from the 1980s), Eleisa Heron, Michael Stuart (80s colleague), Frank Bannister (80s colleague), Cathal Walsh, Peter Craig (Adrian's second PhD student), Brendan Murphy, Antony Unwin (80s colleague), Aideen Keaney, Donal O'Donovan (80s colleague from

Mathematics), Simon Wilson, Adrian, Eamonn Mullins (80s colleague), John Haslett (Adrian's M.Sc. supervisor and 80s colleague), Rozenn Dahyot, Brett Houlding, Olivia Lombard (department administrator in the 80s).

One important thing I did at Trinity was working with John Haslett on a project to evaluate Ireland's wind energy resource. It was a collaborative project with the Irish Ministry of Energy and Met Éireann, the Irish Weather Service. It involved a lot of meteorological data and the development of new methods for time series modeling. The project showed that wind energy was promising for Ireland, but government interest in it dissipated in the late 1980s. However, it has been revived in recent years and more recently Ireland has been producing a lot of wind energy.

During that period at Trinity, several strands of my research started. One was Bayesian model selection, where I got interested in point processes and analyzed a dataset on coal mining disasters. I realized there was a change point, and I developed a Bayesian approach to test and compare models. I also developed a frequentist approach, but the Bayesian approach was simpler and better, which led me to consider Bayesian approaches in later projects.

Another project I worked on was character recognition with Fionn Murtagh, a computer scientist. We used Gaussian distributions centered around lines to detect characters. It was the beginning of the model-based clustering approach that I have been involved with since then.

At the time Trinity College Dublin was a small university, there were maybe four or five hundred faculty and it was very collegial. While there I also started working on social mobility in collaboration with John Jackson, a sociologist. This collaboration got me interested in sociology and led to lifelong collaboration and friendship. John brought me to a conference in Amsterdam in 1984, where I met Mike Hout who was an American sociologist, who happened to be on sabbatical in Dublin that year. Mike and I started collaborating after that.

At Trinity, I developed the mixture transition distribution model, which is a model for discrete value time series that represents higher-order dependence with fewer parameters. The model came out of ideas from my PhD. It was published in JRSSB in 1985, and although it hasn't had a huge impact, it's one of the papers I'm most proud of.

## How did you end up at the University of Washington?

I went on sabbatical to University of Washington (UW) statistics department in 1985. I found that the US as an environment for science was extraordinary. It was at a time when I had no access to email, there was no web, and international travel and even long-distance phoning were extremely expensive. So, I didn't have many contacts in the US, except for some people whom I had written letters to, to ask for copies of their papers that I didn't have access to. Coming to UW, everything was so different. The Science Citation Index (later Web of Science) was there and the library was very comprehensive; all I had to do was go to the library and photocopy the articles that I wanted. People started inviting me to give talks and I'd meet a lot of people and I came to realize that they had actually read my papers. I had been publishing papers but I never got much reaction to them when I was in Ireland.

The UW statistics department was young and dynamic and a terrific place to be. I think it was six years old at the time and the average age of faculty I think was about 35. The department had partly been started by Ted Blalock, who was a sociologist. Blalock was a joint faculty for a few years but then he didn't have enough time to continue doing that. Meanwhile I had been in statistics and I had made contact with the UW sociology department through my social mobility work.  Then they offered me a job that was 50-50 between the statistics and sociology departments. I was excited about it because it was such a terrific environment; the Irish context was also financially difficult for me and my family because the salary was really low. I was very naive about the American system though and asked for tenure immediately, which is what I expected based on my Irish experience, but of course, now, having been through tenure reviews, I realize what a stretch it was. But they came back a few months later and said okay you can have tenure and so then I accepted the position.

*I also remember you saying about the US that you appreciated the typical response to proposing new things?*
Absolutely because in Ireland at the time, when I would occasionally suggest slightly out of the box things and people would say "oh, we've never done that before, that would be difficult" and so on. Whereas in the US, people would say "yeah we've never done that before, let's do it". So, it was the same sentence, but the implication was different. Then they'd say, "Adrian, you can chair the committee", so I learned that part too, but overall, it was for the better.

## How did your research evolve at UW?
Getting the position turned out to be a great piece of luck because it was a fantastic university to be in and a great department. There were several things that I started in terms of research. One area was in environmental statistics; my master's thesis and the wind energy project at Trinity had already involved environmental statistics and UW was the perfect place to continue this line of work because it's one of the strongest universities in environmental sciences. Another long-term project was being involved with the International Whaling Commission. Judy Zeh who was a colleague in statistics was very involved with that and got me involved as well. I also continued working on social mobility with Mike Hout. Also, I continued to work on Bayesian statistics, which led to PhD dissertations by Mike Kahn, Ross Taplin, Michael Newton, Sue Rosenkranz, Danny Walsh, Greg Warnes and Sam Bates Prins.

## Do you want to tell us more about the whaling committee?
I was on the IWC's scientific committee from 1988 to 1998, for 10 years, and I did statistical analysis of whales to inform policy. It was a great experience because whaling was a very controversial area with political implications. It was also enlightening to see the entire problem from data collection to data analysis, through the scientific process, to what actually happened in terms of policy.

The mission of the International Whaling Commission was to strike a balance between whaling and preserving the whale species. It was one of the first big environmental issues; "Save the whale" was a big slogan in the 1970s. It was an interesting problem because whales were viewed as good from an environmental point of view, but Aboriginal peoples were also important and they had been catching whales sustainably; whales were endangered only due to generations of white people catching them. So there was quite a lot of politics involved with it. Some countries, like the United States, the UK, and several other developed countries, believed that commercial whaling shouldn't be allowed, but aboriginal whaling should be allowed. Then there were extreme environmentalists, like Sea Shepherd and some other organizations, and they were part of the Scientific Committee as well, and opposed all whaling, including Aboriginal whaling. Finally there were the whaling countries, which were Japan, Norway, and Iceland, that completely opposed both sets of positions. The Norwegians felt that the US position was that whaling should be forbidden for everybody except US citizens (because Aboriginal whaling allowed whaling by Inuit peoples, Aboriginal people in Alaska, Greenland, and Siberia) but not commercial whalers in Norway. Hence the Norwegians fiercely attacked our work.

*What work did you do exactly and what were these attacks like?*
The commission produced estimates of population dynamics, stocks and trends, and these were used to set quotas to limit the risk to the stock. They defined the acceptable probability of the stock going down as no more than 5%. Prior to our work, the commission primarily used deterministic simulation models for population dynamics and scenarios. However, it was interesting to note that they employed deterministic models while dealing with a probabilistic problem. This discrepancy reflected the division between scientific disciplines that relied on statistical modeling, and those that used deterministic mathematical models based on differential equations. During an IWC scientific committee meeting in 1991, there was a significant moment when the chair confronted us with the challenge of determining the 5% risk threshold. Despite the presence of brilliant scientists from various fields, nobody knew how to accomplish this task. Witnessing this, I recognized it as an extraordinary research opportunity and formulated the Bayesian melding idea.

Some years later, in 1994, the approach was reviewed in a two-week IWC meeting where attacks on our work occurred daily. Judy and I had to stay up overnight, rerunning analyses and preparing responses. If any attacks had stuck, the method wouldn't have been used. In the end, it was accepted because the science was solid. This breakthrough sparked numerous publications and PhDs by Geof Givens and David Poole.

It was also interesting to see that the Inuit people had a good sense of the overall biology and ecology of whales and the Arctic. Initially, biologists made mistakes and greatly underestimated the number of whales. For example, in the 1980s biologists said there were only a few hundred bowhead whales left, but the Inuit people said there were 6,000. Biologists initially dismissed the Inuit people's claim that the whales swim under the ice and breathe through cracks. After further research, it was found to be true, and that there had indeed been around 6,000 whales at that time. It was an interesting revelation. I received a plaque from the Inuit community, thanking me for helping to preserve the Eskimo way of life. It's one of the awards I'm proudest of because it represents something concrete and important.

### Did your research around model-based clustering also take off at UW in the 1990s?

Yes, that's correct. While at the University of Washington, I had the opportunity to collaborate with exceptional PhD students. Fionn Murtagh and I had initiated an approach to clustering using mixture models with geometric constraints. Jeff Banfield, a talented student at UW, took these ideas and expanded them into a rich framework. Collaborating with him was remarkable, as he had an unusual ability to quickly grasp concepts and present innovative solutions. Together, we published the Banfield and Raftery paper in 1993, which played a significant role in advancing geometry-based model-based clustering. Later, Russ Steele, Nema Dean and Derek Stanford did PhDs in the area.

Chris Fraley, who joined the research project in 1991, contributed as a software developer and had a deep understanding of numerical linear algebra. She developed the mclust package, which was remarkably reliable with virtually no bugs; the software was later ported to R by Ron Wehrens. Since Chris retired, the package has been developed by Luca Scrucca, who has helped make it into one of the most downloaded R packages.

This software gained widespread use and led to the formation of the working group on model-based clustering in 1994 (Figure 3, 4). The working group started in 1994 as a weekly gathering at UW and evolved into an annual summer event, where talks were given in the mornings, and afternoons were dedicated to collaborations.

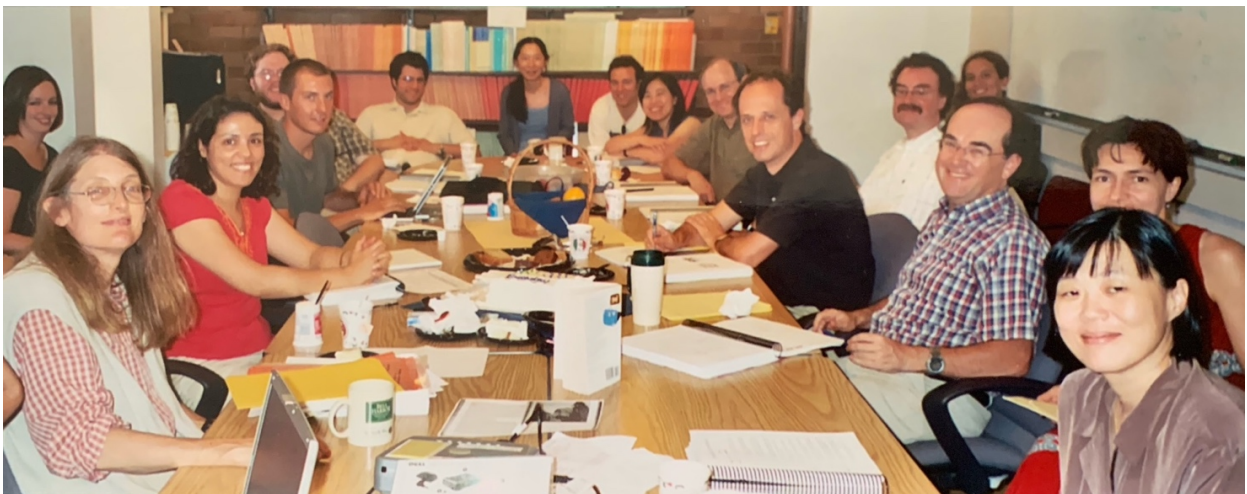

Figure 3: Working group on model-based clustering, Seattle, summer 2004. Left to right: Nema Dean, Chris Fraley, Halima Bensmail, McLean Sloughter, Raphael Gottardo, Russ Steele, Qunhua Li, Matthew Stephens, Ka Yee Yeung, Marc Scott, Denis Allard, Fionn Murtagh, Adrian, Veronica Berrocal, Naisyin Wang, Marina Meila.

The idea of a working group was relatively new in statistics at the time. I initiated it to supervise my three PhD students and Chris, who were part of a collaborative project funded by the Office of Naval Research. The dynamic format of the working group encouraged active participation, lively discussions, and networking. It proved to be incredibly fruitful, leading to numerous ideas and collaborations. This format of regular interaction and collaboration was more common in biology, but now it has become widespread in statistics too.

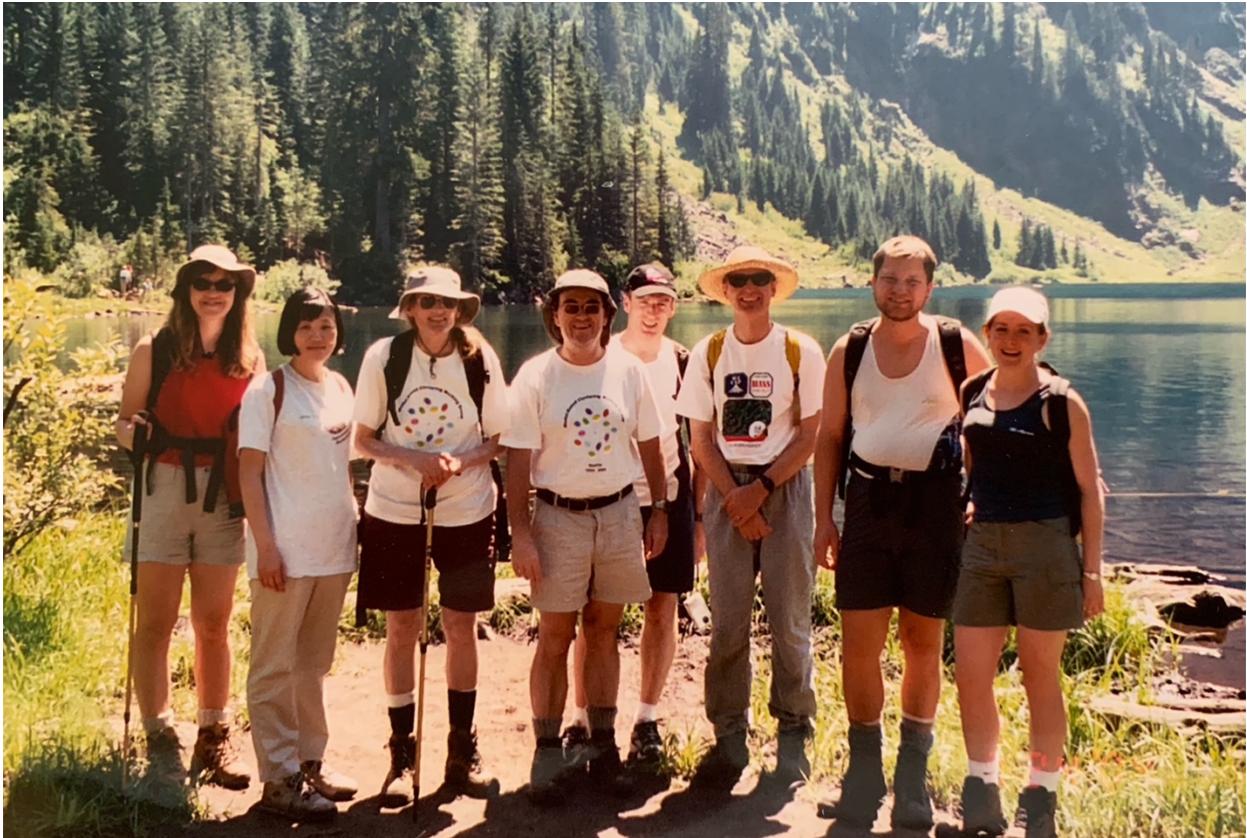

Figure 4: Hiking with coauthors, Washington Cascades, 2004: Hana Ševčíková, Naisyin Wang, Chris Fraley, Adrian, Brendan Murphy, Mark Handcock, Tilmann Gneiting, Claire Gormley.

### What other research areas or projects stood out for you at UW in the 90s?

I did a lot of work on social mobility with Tim Biblarz, who was a sociology PhD student. We looked at social mobility and family structure, analyzing if people do well relative to their social origins based on their family dynamics. You know, intact family or non-intact family with a single mother or single father, these kind of things. It wasn't statistically innovative, and existing methods worked well, but the findings were important, I think. Conventional wisdom was that children always did better in biological two-parent families, but we found that children of working single mothers were as successful.

I also got involved in demographic research, particularly a project with Charlie Hirschman on fertility in Iran. The data were collected during the late 1970s, around the time of the Islamic revolution. The researcher, Akbar Aghajanian, fled Iran with the data and eventually landed in Seattle. We had longitudinal individual-level data, allowing us to address questions about fertility decline and its causes. We developed new methods for analyzing the data and published several papers. I had a brilliant PhD student, Stephen Lewis, who worked on this project as well. He later became a research scientist in the UW School of Social Work.

Bayesian model averaging (BMA) also ties closely with my work on Bayesian model selection, dating back to the 1980s. I developed methods for BMA for generalized linear models and graphical models. A lot of this work was with David Madigan, and we coined the term "model averaging" in a 1994 paper. We co-supervised two Ph.D. students in this area, Jennifer Hoeting and Chris Volinsky. Chris later won the first Netflix Prize, using methods that evolved from his Ph.D. on BMA. In 1995 Rob Kass and I published the paper "Bayes factors", that was later widely cited. A lot of it was actually on BMA, and I regret not including "Bayesian model averaging" in the title, as that part of it didn't get as much attention as I think it should have.

## You had a key role in forming the Center for Statistics in the Social Sciences, CSSS. How did that come about?

There were a lot of precursors to CSSS at UW. It was a sort of perfect storm. I had a position in statistics and sociology, and I think that in the United States there were only two other people who had such positions; Leo Goodman at Chicago and Clifford Clogg at Penn State. But my faculty position was fairly visible at UW and I was getting a lot of requests for help from around the social sciences for statistical help. I asked for a research assistant to help me with these requests and the university gave this to me. The research assistant was Jennifer Hoeting, who later became my PhD student. She was great and did a lot of consulting for social scientists. But after 3 years, the university canceled the position. With requests continuing to come in and me not having enough time to help, quite a movement started in the social sciences at UW, where they were saying, we need statistical help – not just consultation, but also real collaboration.

At the same time the University Initiatives Fund (UIF) had been created at the University of Washington, whose idea was to set aside a certain amount of the university budget to create new interdisciplinary initiatives. So, the idea of a statistics and social sciences interdisciplinary center gained momentum. The selection process for UIF centers was quite formal, and we ended up with 36 faculty or so being on this proposal. It got funded and we started the Center for Statistics and the Social Sciences (CSSS), which was the first center of its kind in the US. It was a very dynamic environment. Key people at UW included Rob Warren, Ross Matsueda and Kate Stovel in sociology, David Madigan and Thomas Richardson in statistics, and Mike Ward in political science. In starting it up, we owed a lot to Steve Feinberg at CMU. He had done a lot of work on statistics for the social sciences and had trained some excellent students in that area. We hired two of them, Elena Erosheva and Adrian Dobra, as some of the first CSSS core faculty members (Figure 5).

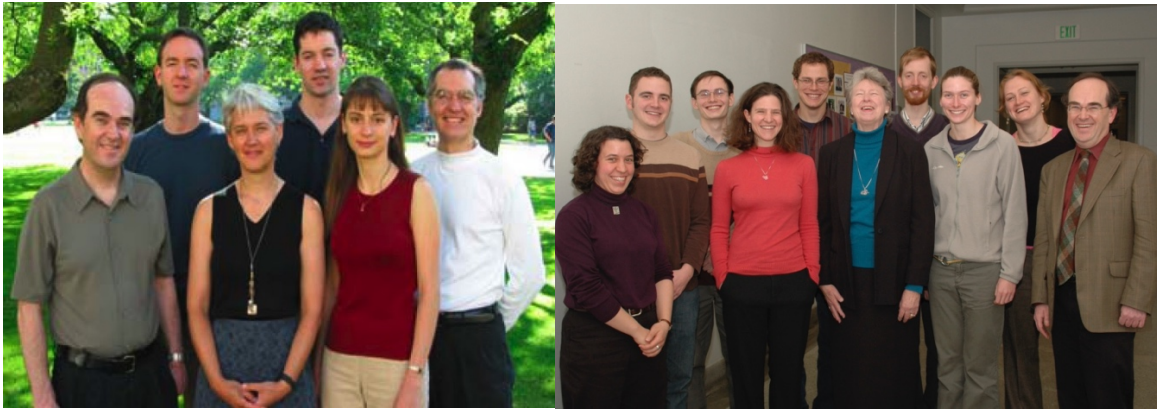

Figure 5: Left: Original CSSS core faculty, early 2000s: Adrian, Peter Hoff, Martina Morris, Kevin Quinn, Elena Erosheva, Mark Handcock. Right: Some of the first CSSS PhD track graduates, mid-2000s: Krista Gile, Tyler Corwin, Richard Callahan, Gail Potter, Jason Thomas, Ann Blalock (widow of Tad Blalock), Thomas Richardson (CSSS associate director), Susan Shortreed, Leontine Alkema, Adrian.

Over time there was a group of 30 to 40 people from a range of disciplines who came to the seminar series and they formed a community that was synthesizing a lot of the skills that were available in statistics and the social sciences.

Intellectually, what I got out of CSSS was getting involved in social networks. We hired as core faculty Peter Hoff and Mark Handcock, who were both involved in social network analysis. After talking to Mark and Peter, I realised that work that I had done on Bayesian multidimensional scaling with Man-Suk Oh, a visitor from Korea, could also work for social networks. The latent space model came out of this work and it has been a great success for social network analysis. Research based on this model remains very active, including in Dublin and France.

## The present
### Moving on to your current projects, what are you working on?
I have focused on bigger scientific projects, particularly since the turn of the century. These include Bayesian model averaging and genetics, weather forecasting, various projects in climate science, and finally, work related to the HIV/AIDS pandemic and demography.

### Let's start with Genetics and Bayesian Model Averaging, how did you get involved in that?
I got involved in research on high-throughput biology (gene expression) data, and similar things in the early 2000s with Ka Yee Yeung, who's a computer scientist, and Roger Bumgarner, who's a microbiologist at UW. We got a grant from NIH to develop BMA methods for assessing genetic risk factors for particular diseases, and finding genetic regulatory networks. BMA turned out to be useful for these because there are, for example, for genetic networks, depending on how you count, about 3000 squared possible links of which only a small number are real scientifically. So shrinkage of some sort is important. Also, BMA turned out to be a way that was quite successful at doing that because it could mobilize other scientific information, which is one of the strengths of the Bayesian approach. Raphael Gottardo and Chad Young worked on this during their PhD studies.

## What did you as a statistician contribute to weather forecasting?

Weather forecasting using numerical weather prediction has been around for about 120 years, and it is mathematically, computationally, and scientifically sophisticated. However, it didn't effectively use statistics until recently. We wanted to improve probabilistic forecasting, which had been done using ensemble forecasting, but hadn't yielded well-calibrated results despite 40 years of effort. To approach it differently, we started a project involving Cliff Mass, a well-known UW meteorologist, Tilman Gneiting, another UW statistician, and a group of psychologists led by Buz Hunt and Susan Joslyn. One aspect we focused on was how best to communicate uncertainty, realizing that it's more of a cognitive problem than a statistical one. Collaborating with cognitive psychologists was enlightening as we discovered that our preconceived ideas didn't align with cognitive reality. We developed statistical and cognitive methods and websites, and received substantial funding, including $5 million from the Office of Naval Research. Our work had a significant impact worldwide, leading to the adoption of probabilistic weather forecasting methods by various agencies. Several PhD students made key contributions to the research, including Veronica Berrocal, McLean Sloughter and Will Kleiber.

## You also mentioned climate science, can you tell us more about that?

The biggest project in climate science was with Cecilia Bitz, a prominent climate scientist at UW, on sea ice forecasting. This area lacked involvement from statisticians, and the physics was still in early stages. However, the National Science Foundation had identified it as part of one of the major scientific challenges for the future. We made some statistical contributions. Hannah Director did her PhD on this project, and we also explored interesting mathematical aspects, such as the analysis of star-shaped spatial sets.

Another project was on probabilistic climate change forecasting with Dargan Frierson, a professor of atmospheric science, and Dick Startz, an economist who was then at UW. This again was an area that has mostly been deterministic. The IPCC relies a lot on scenarios, whose interpretation is a bit vague. So we developed a statistical model that gave a probabilistic forecast of climate change. It was quite well received, including in the media, because people seemed to find it easier to understand "there's a 95% probability that the climate change on current trends will be greater than two degrees" than "there are 40 different models, and they say different things, but mostly they seem to say go in this direction". At UW, they made me an adjunct professor of atmospheric science, which I was really proud of because I think it's a really terrific department; it was actually just ranked number one in the United States.

Our work on climate change also led to work on assessing the social cost of carbon, defined as the social cost (in dollars) of putting an extra ton of carbon into the atmosphere. By law, that number needs to be taken account of by the US government in decision making. Initially the social cost of carbon was set at $51 but that was calculated in a fairly ad hoc way. The National Academy of Sciences called for better methods, including probabilistic population and economic

forecasting. So we got involved in that, and our group has been working with several Federal agencies to do that. We published some papers which use probabilistic ideas and estimates the social cost of carbon to be a lot higher, about $185. This could make quite a difference to practical climate policy and it's making its way through the administration now to actually be implemented.

**Last but not least, you have done a lot of work in the areas of HIV/AIDS and probability population projections with the UN, how did those projects come about and what are some notable outputs?**

My work on HIV/AIDS was with the UNAIDS reference group, which deals with the modeling for the UNAIDS, the organization that leads the response to the HIV/AIDS pandemic. UNAIDS had mostly relied on deterministic models for their estimation and forecasting, using epidemiological infectious disease models, and there was a big need for both improved estimation of model parameters and also for assessing uncertainty, because there is a lot of uncertainty in these things. I was able to propose the Bayesian melding approach, which had been first developed for the whales and turned out to work very well. You pioneered that, Leontine, and Le Bao was a PhD student who played a leading role on that project; he's now a professor at Penn State. A lot of that work was also with Sam Clark, who was a UW demographer at the time, and has worked a lot on infectious disease epidemiology. My involvement with that project was from 2008 to 2015 and I felt it was fairly complete. Le Bao has continued working with the UNAIDS reference group and has become a national leader in this area.

The project on probabilistic population projections with the UN was initiated by the UN. I was contacted by Thomas Buettner, who was initially the Chief of the Population estimates and projections section and, later on, the Assistant Director of the UN Population Division. The UN Population Division at that time was using traditional deterministic approaches to population forecasting and estimation. Thomas was interested in seeing if they could do better using modern statistical methods. He had asked a number of demographers, and they were aware of my work on whales. So Thomas contacted me to discuss using modern methods for humans instead of whales.

The rest is history, we got started. You, Leontine, know that part well because you went to work at the Division as a PhD student intern very early on. That was important to create the link between our group and the UN, and also to define the statistical and technical issues. I think it's the only time that a PhD student has really been involved in the very early stages of a project. We got an NIH grant to work on probabilistic projections that started in 2006. In 2015, the UN adopted the methods officially as the basis for their official population forecast after quite a few years of testing them out and expert critique and evaluation and so on. The UN Population Division is an amazing organization because even though it is a large official organization that could be bureaucratic, the people working in the division are extremely committed to improving methods and producing high quality estimate and projections. Patrick Gerland, who is currently the Chief of the Population estimates and projections section, has been a key person in terms of making changes. Successive directors Hania Zlotnik and John Wilmoth have also been very supportive and committed to innovating and improving methods.

More broadly, I think that this project has had a big impact on demography, in the sense that Bayesian methods are now much more commonly used. The project has also been very fruitful in terms of PhD students. Five PhDs graduated with dissertation research that was focused in this area: Leontine, Mark Wheldon, who developed Bayesian population reconstruction, Jon Azose, who worked on migration, Yicheng Li, who worked on smoking and mortality, and Peiran Liu, who worked on accounting for uncertainty in historical measurements. There are also PhD students in progress. My wife, Hana Ševčíková, has also played a major role. She is a research scientist producing software and has also made important research contributions to the demography project. She produced all the software, which has been critical to its widespread adoption. Overall, this has probably been my biggest project over the past 15 years.

### In addition to your research and University responsibilities, you seem to have been very active in the National Academy?

Yes, I got elected to the National Academy of Science in 2009 (Figure 6, 7). And it's been very rewarding because the mission of the National Academy of Science is not just to elect and honor people, but to advise the US federal government. They issue one report every working day on average, providing recommendations to different government agencies. It's a very active organization.

I've also been on the editorial board of PNAS, which is one of the top general science journals. They are trying to include more papers from underrepresented disciplines like the social and physical sciences so I've been encouraging people in sociology, statistics, and demography to submit more papers. Over the past 10 years, we have seen an increase in the number of papers from these disciplines, which I believe is a positive opportunity for these disciplines because they can have a significant impact. I've also been involved in setting up a statistical review committee for PNAS to help improve the quality of statistical analyses in the journal.

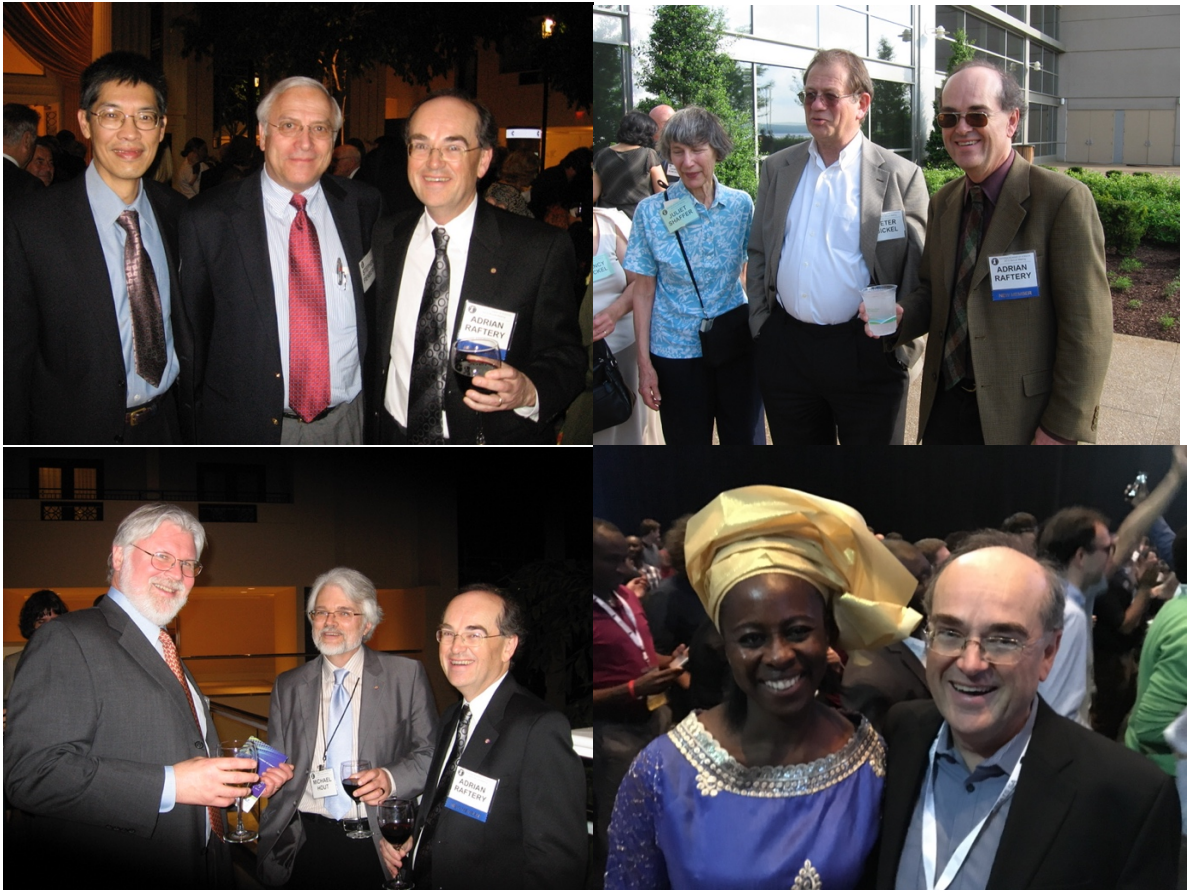

Figure 6: With senior colleagues in the 2010s. Clockwise from top left: Wing Wong and Steve Fienberg; Juliet Shaffer and Peter Bickel; demographer Anastasia Gage (President of the International Union for the Scientific Study of Population); sociologists Doug Massey and Mike Hout.

## What are you most proud of in your career?

It's hard to say - scientists often struggle to judge their own contributions. I don't have a particular project or accomplishment that stands out as my proudest moment. Demography and atmospheric science have probably had the most real-world impact. However, if we look at citations, my work on Bayesian model averaging, Bayesian model selection, and model-based clustering has also had significant impact. But it's important to note that these methodologies have had more time to accumulate citations.

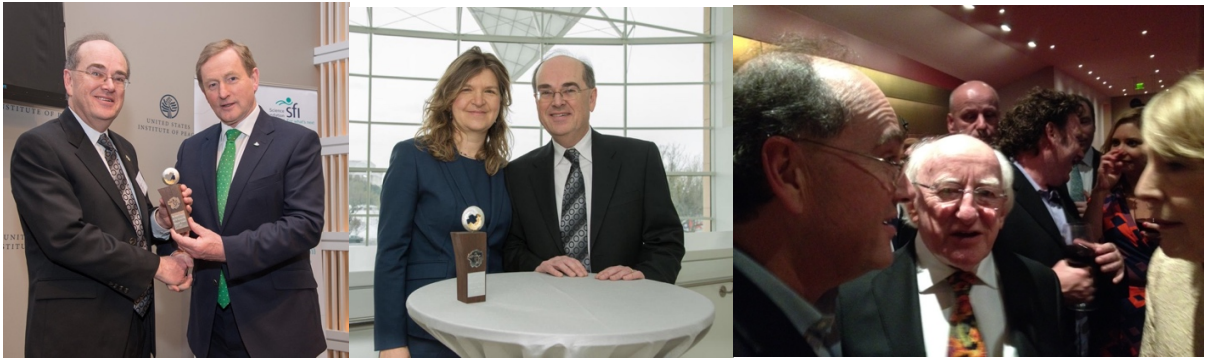

Figure 7: Continuing links with Ireland. Left: Adrian getting the St. Patrick's Day Medal from Irish Taoiseach (Prime Minister), Enda Kenny, 2017. Center: With his wife, Hana Ševčíková, at the ceremony. Right: With Irish President Michael D. Higgins and his wife, Sabina Higgins.

## You have two children, have they followed in your statistician's footsteps?

No, my children have been more involved in qualitative than quantitative fields, although they have added in a little statistics later in their careers. My daughter is a journalist; she has taken statistics classes and organized statistics training for journalists. My son, who initially had no interest in statistics, later found it useful in his work on analytics in the tech sector (Figure 8).

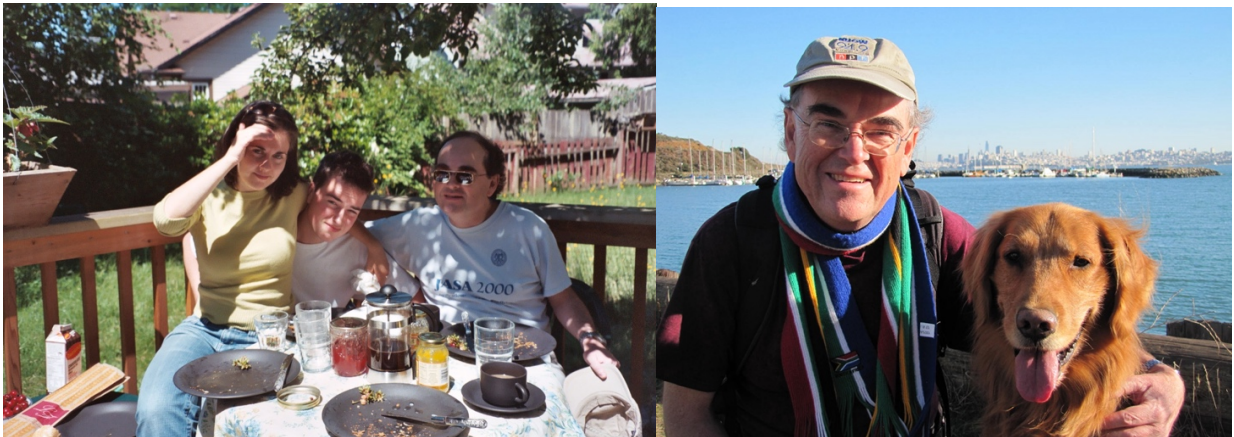

Figure 8: Adrian with his children, Isolde and Finn, in 2005, and with his dog, Jasper, in 2017.

## Let's end with the future, what do you see for the discipline?

I believe the future is very bright for the discipline! Statistics has always been about developing new methods for scientific, technological, and policy problems in collaboration with other fields. This interdisciplinary aspect has been a defining characteristic since its early days. There is a constant demand for statistical methods in various domains, as we have seen during the COVID pandemic. I think the future of methodology will be driven by emerging data challenges and the needs of different scientific disciplines. Areas like gene expression data, demography, atmospheric sciences, and infectious disease epidemiology continue to offer promising opportunities for statisticians.

My advice would be for statisticians to seek out new areas that require innovative statistical methods and contribute to solving important scientific problems. Statistics has been at the forefront of scientific progress for over a century, developing methods that drive discoveries and breakthroughs. I believe it will continue to be a vital discipline, adapting to new data challenges and contributing to various scientific fields.

## That's an inspiring outlook for the future of statistics. Is there anything else you'd like to share?

Overall, I feel fortunate to have grown up where I did, to have studied statistics and worked at UW. It's been a great collaborative environment, and I've coauthored with 10 different faculty members in the UW statistics department alone. UW has also been supportive of collaborations with sociologists and atmospheric scientists. I've had exceptional PhD students who have been a real joy to work with.

The United States, in general, has provided a great environment for scientific research over the past several decades. I feel fortunate to have had many outside collaborators and to have worked in supportive environments throughout my career. I realize how privileged I have been to work in resource-rich environments. Research can be incredibly challenging in resource-poor settings, and it's important to acknowledge the magnitude of our privilege as researchers in wealthy countries.

**Acknowledgements:** This work was supported by NIH grant R01 HD-070936 and Science Foundation Ireland Insight Research Centre grant SFI/12/RC/2289_P2.